\documentclass[12pt]{iopart}
\usepackage{iopams}

\begin{document} 

\title{The role of initial conditions in the ageing of the long-range spherical model} 
\author{Sreedhar B. Dutta}
\address{Korea Institute for Advanced Study, 87 Hoegiro, Dongdaemun-gu, Seoul 130-722, 
South Korea}
\begin{abstract}
The kinetics of the long-range spherical model evolving from various initial states 
is studied. In particular, the large-time auto-correlation and auto-response functions 
are obtained, for classes of long-range correlated initial states, and for magnetized 
initial states. The ageing exponents can depend on certain qualitative features of 
initial states. We explicitly find the conditions for the system to cross over from 
ageing classes that depend on initial conditions to those that do not.    
\end{abstract}

\section{Introduction}
\label{intro}
Slowly-relaxing thermodynamic systems are being intensely studied for decades now, and 
for numerous reasons. One of the main objectives is to understand genuine 
non-equilibrium states: effecting their description, finding and categorizing
the large-scale properties, identifying the necessary qualitative features of 
each universality class, determining sufficient exponents that would enable us to label 
the classes, and answering a host of other similar questions usually posed in 
equilibrium statistical physics.       
  
Systems that exhibit phase transitions relax slowly and may not equilibriate
when either quenched to critical temperature or below. For critical quench 
the relaxation is inhibited due to long-range correlations \cite{hohenberg}, while when 
quenched from above to below the critical temperature, it is due to the competition 
between locally-equilibriated domains for a global equilibrium state \cite{bray94}. In 
either of these cases, after waiting for a long time $t_w$, much longer than the 
transient microscopic time-scales, these systems which are not in equilibrium are 
expected to become statistically scale invariant, when scaled by a characteristic 
time-dependent length-scale $L(t)$. The length grows in time algebrically, 
$L(t) \sim t^{1/z}$, defining the dynamic exponent $z$. The scaling hypothesis, if 
applicable, proposes that the two-point functions of observables, say correlation- or 
response- function, take the form 
\begin{eqnarray}
\label{two-pt-fn}
\left\langle {\cal O}_i(\rm{x},t) {\cal O}_j(\rm{y}, t_w) \right\rangle = 
t_w^{-a_{ij}} F_{ij}\left(\frac{|\rm{x}-\rm{y}|}{t_w^{1/z}}, \frac{t}{t_w} \right),
\end{eqnarray}
for large observation time $t > t_w$; where ${\cal O}_i(\rm{x},t)$ refers to some local 
observable at time $t$ and position $\rm{x}$, $a_{ij}$ is a universal exponent, 
$F_{ij}$ is a scaling function. 
Here it is assumed that the system is translational- and rotational- invariant. 
Furthermore, in the ageing regime, namely when $t \gg t_w$, the asymptotic behavior of 
the function $F_{ij}$ is as follows, 
\begin{equation}
\label{two-pt-fn-2}
F_{ij}\left(0, t/t_w \right) \sim (t/t_w)^{-\lambda_{ij}/z} ,
\end{equation}
and defines an ageing exponent $\lambda_{ij}$. The exponent $a_{ij}$ though may not be 
an independent exponent. For instance, when $t \approx t_w$, and 
$|\rm{x}-\rm{y}| \ll L(t)$, and further under the conditions that would establish 
equilibrium correlations of size $L(t)$ in time $t$, the two-point function of 
observables at critical quench would behave as
$\left\langle {\cal O}_i(\rm{x},t) {\cal O}_j(\rm{y}, t_w) \right\rangle 
\sim |\rm{x}-\rm{y}|^{-\Delta_{ij}} $, 
where $\Delta_{ij}$ is an equilibrium exponent. If this is so for 
(nearly) equal-time observables, then consistency with scaling hypothesis would 
imply $a_{ij} = \Delta_{ij}/z$, and hence is not an independent exponent.

Scaling hypothesis holds in many slowly-relaxing systems, and is tested experimentally
and theoretically\cite{calabrese05}. Long-range spherical model is one of the few 
models that can be solved exactly. This model when evolved from an uncorrelated 
initial state shows scaling behaviour at large times\cite{florian}. Initial conditions 
are known to affect the large-scale properties in some short-range models\cite{janssen, 
bray91, ritschel, picone, fedorenko, annibale06, calabrese06, calabrese07, annibale08}. 
So it would be interesting to know how, in general, the initial conditions influence the 
scaling behaviour, and in particular, that of long-range models.  
It is this issue that we address in this paper, after having solved the non-conserved 
long-range spherical model evolving from different initial states.
We evaluate the correlation and the response functions for various initial conditions, 
and determine their scaling behaviour, and also find how the ageing exponents depend on 
certain features of the initial state. In some cases the $a_{ij}$ exponents turn out to 
be independent, implying that the spacial correlations within regions of size $L(t)$ 
after a time $t$ are not governed by canonical equilibrium fluctuations.

The layout of the paper is as follows. In the next section, the model is defined and 
an outline of the formal solution is provided. In section \ref{sec:LR-ini}, the 
scaling behaviour of auto-correlation and auto-response functions, with long-range 
initial correlations is investigated. In section \ref{sec:mag-ini}, the model is 
analysed with an initial state that has no correlations but has non-vanishing 
magnetization. Section \ref{conc} concludes with a brief summary and a few remarks.

\section{Long-range spherical model}
\label{sec:model}
A spherical model is exactly solvable not only in equilibrium but also when evolving 
with relaxational dynamics\cite{Godreche00}. The techniques used in solving that model 
can be easily extended to a long-range spherical model\cite{florian}. In this section, we 
first define the long-range model and then solve for the correlation and response functions.

\subsection{The model}
The equilibrium long-range spherical model \cite{Joyce66} is described by the 
Hamiltonian
\begin{equation}
\label{hamiltonian}
{\cal H} = -\frac{1}{2}\sum_{\rm{x},\rm{x}'} J_{\rm{x},\rm{x}'}S_{\rm{x}}S_{\rm{x}'} ,
\end{equation}
where the coupling between the spins is long range and given by
\begin{equation}
\label{coup}
J_{\rm{x},0} = J_c~|\rm{x}|^{-(d+\sigma)},
\end{equation}
such that $\sum_{\rm{x}} J_{\rm{x},0}=1$, which in turn fixes the constant $J_c$.
The variable $S_{\rm{x}}$ respects the spherical constraint $\widehat{Y}[S]=0$, where 
\begin{equation}
\label{Y}
\widehat{Y}[S] = \frac{1}{N}\sum_{\rm{x}\in\Lambda} S_{\rm{x}}(t)^2  - 1 ,
\end{equation}
and, $\rm{x}$ runs over the sites of the $d$-dimensional lattice $\Lambda$, 
and $N$ is the total number of sites on the lattice.

The coarse-grain dynamics near the critical temperature or the coarsening 
dynamics when quenched from above to below $T_c$, for the non-conserved order-parameter 
$S_{\rm{x}}$ is given by following Langevin equation, 
\begin{equation}
\label{langevin}
\partial_t S_{\rm{x}}(t) =
- \left. \frac {\delta {\cal H} } {\delta S_{\rm{x}} } \right|_{S_{\rm{x}}
\rightarrow S_{\rm{x}}(t) } - S_{\rm{x}}(t) \widehat{Z}(t)  + \eta_{\rm{x}}(t) ,
\end{equation}
\begin{equation}
\label{noise-corr:R}
\left\langle \eta_{\rm{x}}(t)\eta_{\rm{x}'}(t')\right\rangle = 
2T \delta(t-t')\delta_{\rm{x},\rm{x}'} ~,
\end{equation}
where $\widehat{Z}$ is the Langrange multiplier that is determined by the 
constraint. Assigning Stratonovich convention to the stochastic dynamics will 
lead the constraint to the expression,
\begin{equation}
\label{Z}
\widehat{Z}(t) = -\frac{1}{N} \left(2 {\cal H}(t) 
- \sum_{\rm{x}} S_{\rm{x}}(t)\eta_{\rm{x}}(t)\right)
\end{equation}

\subsection{Formal solution}

We will first separate the Gaussian and non-Gaussian terms in equation (\ref{langevin})
so as to solve. We will see that the fluctuations of the Lagrange multiplier can be 
neglected when restricted to local observables in the limit $N \rightarrow \infty$. 
Furthermore, the spherical constraint can be replaced by a mean-spherical constraint in 
this thermodynamic limit. We will then explicitly write down the solution in the 
Gaussian regime using this mean-spherical constraint.   
 
\subsubsection{Fluctuations of the Lagrange multiplier} 
~\\
The fluctuating part of $\widehat{Z}(t)$ is of $O\left(1/\sqrt{N}\right)$, and 
hence $S_{\rm{x}}$ can be written as,
\begin{equation}
\label{decomp}
S_{\rm{x}}(t) = s_{\rm{x}}(t) + \frac{1}{\sqrt{N}} \xi_{\rm{x}}(t),
\end{equation}
where $s_{\rm{x}}$ captures only the Gaussian fluctuations while $\xi_{\rm{x}}$ accounts 
for the remaining non-Gaussian fluctuations. 

Within the Gaussian approximation the mean-square fluctuations in 
$\widehat{Y}[s]$ can be estimated from the following relation, obtained using 
Wick's theorem and translation invariance:
\begin{equation}
\label{Y-fluc}
\langle \widehat{Y}^2[s] \rangle - \langle \widehat{Y}[s] \rangle^2
\approx \frac{2}{N} \sum_{\rm{x}} C_{\rm{x},0}^2(t,t),
\end{equation}
where $C_{\rm{x},0}(t,t)$ is the equal-time spin-spin correlation function. 
In the case of uncorrelated initial state, the late-time behaviour of the 
correlation function is as given below\cite{florian},
\begin{equation}
\label{Cor}
C_{\rm{x},0}(t,t) \sim \left\{ \begin{array}{cc}
G\left(|\rm{x}|t^{-1/\sigma}\right),~ & T < T_c  \\
t^{1-d/\sigma}G\left(|\rm{x}|t^{-1/\sigma}\right),~ &  T = T_c  \end{array} \right. ,
\end{equation}
where, $G(\rm{x}) \propto \int_k \exp\left(i k \cdot \rm{x} - 2|k|^{\sigma}\right) $.
Hence we obtain,
\begin{equation}
\label{Y-fluc2}
\langle \widehat{Y}^2[s] \rangle - \langle \widehat{Y}[s] \rangle^2
\propto \left\{ \begin{array}{cc}
t^{d/\sigma}/N,~ & T < T_c  \\
t^{2-d/\sigma}/N,~ &  T = T_c  \end{array} \right. .
\end{equation}
Therefore, in the limit $N \rightarrow \infty$ these fluctuations become 
negligible for late-times $t < t^{*} \sim N^{\sigma/d}$ \cite{fusco},
both at and below the critical temperature $T_c$.  
Even at $T_c$ the threshold time, $t^{*}$, is unchanged 
since the phase transition occurs only for $0< \sigma < d$. Thus, 
taking first the $N \rightarrow \infty$ limit is a necessary condition to have 
the same asymptotic behaviour for both spherical and mean-spherical models in the 
Gaussian regime. 

Substituting the decomposition of spin variables, as given in equation 
(\ref{decomp}), into equation (\ref{Y}) and keeping only the 
leading subdominant terms yields the expression
\begin{equation}
\label{Y-fluc3}
\widehat{Y}[S] \approx \langle \widehat{Y}[s] \rangle
+ \frac{1}{N}  \left[ \sum_{\rm{x}} \left( s_{\rm{x}}^2 - \langle s_{\rm{x}}^2 \rangle \right)
+ \frac{2}{\sqrt{N}} \sum_{\rm{x}} s_{\rm{x}} \xi_{\rm{x}}
\right].
\end{equation}
Here the non-Gaussian term is of the same order in $N$ as the Gaussian 
fluctuations, and hence can play a role in certain global observables.

\subsubsection{Solution in the Gaussian regime}
~\\
The Gaussian approximation of the Langevin equation (\ref{langevin}) 
in the Fourier space is given as 
\begin{equation}
\label{EOM}
\partial_t s_k(t) = - \Big( \omega_k + Z(t) \Big) s_k(t) + \eta_k(t) ,
\end{equation}
where $\omega_k$ is the Fourier transform of $-J_{x,0}$, and the noise $\eta_k(t)$ has the variance,
\begin{equation}
\label{noise-corr}
\left\langle \eta_{k}(t) {\eta}_{k'}(t')\right\rangle =
2 T \delta(t-t') N \delta_{k+k'}.
\end{equation}
In the continuum limit, 
$N\delta_{k}$ gets replaced by $(2\pi)^d\delta(k)$, and 
$N^{-1}\sum_k$ is replaced by $(2\pi)^{-d}\int d^dk$.
In this limit, for small $k$, the function
$\omega_k \rightarrow \widetilde{c}_0+ \widetilde{c}_1 |k|^{\sigma} 
+ \widetilde{c}_2 |k|^2 + O(|k|^4)$, where 
$\widetilde{c}_n$ are lattice-dependent constants. We shall restrict
$\sigma < 2$, for $\sigma \ge 2$ the relevant model is the usual 
short-range spherical model. Further, without any loss of generality we shall 
set $\widetilde{c}_0=0$ by absorbing it into $Z(t)$.
 
The solution of the above equation is 
\begin{equation}
\label{solution}
s_k(t) =  \frac {e^{-\omega_k t} } {\sqrt{g(t)}}
\left[ s_k(0) +
\int_0^t d\tau e^ {\omega_k \tau } \sqrt{g(\tau)} \eta_k(\tau) \right] ,
\end{equation}
where the yet to be determined constraint function $g(t) = \exp(2\int_0^t d\tau 
Z(\tau))$.
It is often convenient to write the above equation as
\begin{equation}
\label{solution-2}
s_k(t) =  R_k(t,0)s_k(0)
+ \int_0^{t} d\tau R_k(t,\tau) \eta_k(\tau),
\end{equation}
where Green's function $R_k(t,t')$ satisfies the equation, 
\begin{equation}
\label{Green}
\Big( \partial_t  + \omega_k + Z(t) \Big) R_k(t,t') = \delta(t-t'),
\end{equation}
with the initial condition $R_k(t,t')=0$ for $t < t'$.
It is easy to see that $R_k(t,t')$ is also the Fourier transform of the 
response function $R_{\rm{x},0}(t,t')$, and is explicitly given as
\begin{equation}
\label{res:k}
R_k(t,t') = e^{-\omega_k (t-t')}\sqrt{\frac{g(t')}{g(t)}} \theta(t-t'),
\end{equation}
where $\theta(t)$ is the step function.

If the initial conditions for the spins $s_{\rm{x}}$ are chosen to be translation 
invariant with a variance, 
\begin{equation}
\label{ini-var}
\langle s_{k}(0)s_{k'}(0)\rangle = N \delta_{k+k'} C_k(0,0).
\end{equation}
then, using equation (\ref{solution}), we obtain the correlation function
\begin{equation}
\label{2tcorr:k}
\langle s_{k}(t)s_{k'}(t')\rangle = N \delta_{k+k'} C_k(t,t'),
\end{equation}
where
\begin{equation}
\label{corr:k}
C_k(t,t') = \frac{g(t_m)}{\sqrt{g(t)g(t')}} e^{-\omega_k|t-t'|} 
C_k(t_m,t_m),
\end{equation}
and $t_m = \mbox{min}(t,t')$, and the equal-time correlation function 
\begin{equation}
\label{1tcorr:k}
C_k(t,t) = \frac{e^{-2 \omega_k t}}{g(t)} 
\left[ C_k(0,0) + 2T \int_0^t d\tau e^{2 \omega_k \tau} g(\tau) \right] .
\end{equation}
Note that $C_k(t,t)$ is also the Fourier transform of the correlation function  
$C_{\rm{x}}(t,t) = \langle S_{\rm{x}+\rm{y}}(t)S_{\rm{x}}(t)\rangle$. We now impose the 
mean-spherical constraint, $\langle \widehat{Y}[s]\rangle =0$, which then implies that 
the following condition should hold at all times, 
\begin{equation}
\label{const:k}
\frac{1}{N}\sum_k C_k(t,t) =1 .
\end{equation}
This condition not only fixes $g(t)$ but also restricts the choice of initial 
conditions. Substituting expression (\ref{1tcorr:k}) in the above equation 
gives $g(t)$ to be the solution of the following Volterra equation,
\begin{equation}
\label{g-fn}
g(t) = A(t) + 2T \int_0^t d\tau f(t-\tau) g(\tau),
\end{equation}
with $g(0)=1$, where
\begin{equation}
\label{A-fn}
A(t)= \frac{1}{N}\sum_k e^{-2 \omega_k t} C_k(0,0),
\end{equation}
and
\begin{equation}
\label{f-fn}
f(t)= \frac{1}{N}\sum_k e^{-2 \omega_k t} .
\end{equation}
Note that $A(0)=1$, due to spherical constraint. The function $A(t)$ depends on the 
initial correlations, and hence $g(t)$. We shall solve equation (\ref{g-fn}) 
and determine $g(t)$ for various initial conditions in latter sections, and 
explicitly evaluate the correlation and response functions. 

The non-Gaussian variable $\xi_{\rm{x}}(t)$ can be completely expressed in terms of 
Gaussian observables as has been done recently for the short-range spherical model 
\cite{annibale06}. But we shall not express this here, since we are interested  
in the ageing properties of local observables, for which the non-Gaussian 
corrections vanish in the limit $N \rightarrow \infty$.  
 
\section{Long-range correlated initial conditions}
\label{sec:LR-ini}
In this section, we will evaluate the auto-correlation and auto-response functions, when
evolved from a long-range-correlated initial state.
To this end, we will first calculate the constraint function. 

\subsection{Constraint function}
The large-time solution of the equation (\ref{g-fn}) for the constraint function 
$g(t)$ not only depends on the temperature $T$ but also on the form of the 
initial correlation function $C_k(0,0)$. The case with uncorrelated initial 
conditions in this model was studied recently\cite{florian}. 
We now investigate the behaviour when evolved from a class of initial states with long-range 
spin-spin correlations. In other words, we choose $C_k(0,0)$ such that its small $k$ 
behaviour is given by
\begin{equation}
\label{ini-corr}
C_k(0,0) \approx c_0~|k|^{\sigma_0},
\end{equation}
and also satisfies equation (\ref{const:k}).
In the real space, the large-distance initial correlations become
$C_{\rm{x}}(0,0) \sim c'_0~ |\rm{x}|^{-(d+\sigma_0)}$. The exponent $\sigma_0$ is 
restricted by the condition $d+\sigma_0 > 0$,
on the grounds that the correlations should decrease with increase in distance.
Short-range spherical model with similar initial state was also studied 
recently\cite{picone}.

The Laplace transform of equation (\ref{g-fn}) results in the expression,
\begin{equation}
\label{g-L}
g_L(p) = \frac{A_L(p)} {1- 2T f_L(p)},
\end{equation}
where, $g_L(p)$, $A_L(p)$ and $f_L(p)$ are the Laplace transforms of 
$g(t)$, $A(t)$ and $f(t)$, 
respectively. The small $p$ behaviour of $g_L(p)$ is sufficient for deducing the large-$t$ properties of $g(t)$. The function $f_L(p)$, for small $p$, is given as 
\begin{equation}
\label{f-L}
f_L(p) = - A_0 p^{-1+d/\sigma} + \sum_{n=0}^{\infty} A_{n+1} (-p)^n
\end{equation}
where $A_0=|\Gamma(1-d/\sigma)|\widetilde{A}_0$, and 
$A_n= N^{-1}\sum_k (2\omega_k)^{-n} - \int_k (2\widetilde{c}_1 
|k|^{\sigma})^{-n}$,
and 
$\widetilde{A}_n = \int_k \exp(-2\widetilde{c}_1 |k|^{\sigma})
|c_0 k|^{n\sigma_0}$.
Similarly, $A_L(p)$ is given by
\begin{equation}
\label{A-L}
A_L(p) = - B_0 p^{-1+(d+\sigma_0)/\sigma} + \sum_{n=0}^{\infty} B_{n+1} (-p)^n
\end{equation}
where, $B_0=|\Gamma(1-(d+\sigma_0)/\sigma)| c_0 \widetilde{A}_1$,
and  
$B_n= N^{-1}\sum_k (2\omega_k)^{-n}C_k(0,0) - 
\int_k (2 \widetilde{c}_1 |k|^{\sigma})^{-n} c_0|k|^{\sigma_0}$.

There is a phase transition at temperature $T_c = 1/2A_1$.  
At the critical temperature $T_c$, the small $p$ behaviour of $g_L(p)$ can 
easily be obtained by substituting equations (\ref{f-L}) and (\ref{A-L}) into 
(\ref{g-L}). In this limit, $g_L(p) \approx g_0 \Gamma(\psi+1)p^{-(\psi+1)}$,   
and hence the large-time form of $g(t)$ is given by 
\begin{equation}
\label{g:T=Tc}
g(t) \approx g_0(\sigma,\sigma_0,d) ~t^{\psi(\sigma,\sigma_0,d)} ,
\end{equation}
where the constant $g_0(\sigma,\sigma_0,d)$ and the exponent
$\psi(\sigma,\sigma_0,d)$ depend on the values of $\sigma$ and $\sigma_0$, for a 
given $d$. There are four regimes, as specified in the table(\ref{table1}),  
each distinguished by the form of $\psi$. Though the form of $g(t)$ in case-IV 
and case-V is the same, they are listed separately for, as we shall see 
later, they get distinguished by ageing exponents and the 
fluctuation-dissipation ratio.

For $T<T_c$, in the small $p$ region the function 
$g_L(p) \approx (1-T/T_c)^{-1} A_L(p)$, and hence
\begin{equation}
\label{g:T<Tc}
g(t) \approx  \widetilde{A}_1 \left(1-\frac{T}{T_c}\right)^{-1} 
t^{-(d+\sigma_0)/\sigma} .
\end{equation}

\begin{table}[t]
\[
\begin{array}{||c||c|c|c||} \hline \hline
 \mbox{Regime} & \mbox{Conditions} & g_0(\sigma,\sigma_0,d) & 
\psi(\sigma,\sigma_0,d)\\
  \hline \hline
\mbox{I}    &  d/2 < \sigma < d , & 
\frac{A_1 B_0}{A_0 \Gamma\left(-\sigma_0/\sigma\right)} 
& -1-\sigma_0/\sigma \\
 &  -d < \sigma_0 < \sigma -d & & \\
  \hline
\mbox{II}    &  0 < \sigma < d/2 , & 
\frac{A_1 B_0}{A_2 \Gamma\left(2-(d+\sigma_0)/\sigma\right)} 
& 1-(d+\sigma_0)/\sigma \\
 & -d < \sigma_0 < \sigma -d & & \\
  \hline
\mbox{III}    &  d/2 < \sigma < d , & 
\frac{A_1 B_1}{A_0 \Gamma\left(-1+d/\sigma\right)} 
& -2 + d/\sigma \\
 &  \sigma_0 > \sigma -d & & \\
  \hline
\mbox{IV}    &  0 < \sigma < d/2 , & 
\frac{A_1 B_1}{A_2 } 
& 0 \\
 &  \sigma_0 > \sigma -d ,~ \sigma_0 > -\sigma & & \\
  \hline
\mbox{V}    &  0 < \sigma < d/2 , & 
\frac{A_1 B_1}{A_2 } 
& 0 \\
 &  \sigma_0 > \sigma -d,~ \sigma_0 < -\sigma & & \\
  \hline \hline
  \end{array}
  \]
  \caption{\label{table1} Values of $g_0$ and $\psi$, as defined in equation 
(\ref{g:T=Tc}), characterizing the asymptotics of $g(t)$ in various critical regimes.
  }
  \end{table}

\subsection{Auto-correlation and auto-response functions}
We first write the auto-correlation and auto-response functions in terms of $g(t)$, 
$A(t)$ and $f(t)$, and then analyse their late-time behaviour. The correlation 
function is given by
\begin{eqnarray}
\label{corr-def}
C_{\rm{x}}(t,t_w) := \langle S_{\rm{x}+\rm{y}}(t) S_{\rm{y}}(t_w)\rangle 
= \frac{1}{N} \sum_k e^{i k \cdot \rm{x}} C_k(t,t_w), 
\end{eqnarray}
and hence the expression for the auto-correlation function, 
$C(t,t_w):= C_{\rm{x}=0}(t,t_w)$,
using equations (\ref{corr:k}) and (\ref{1tcorr:k}), becomes
\begin{equation}
\label{autocorr}
C(t,t_w) = \frac{1}{\sqrt{g(t)g(t_w)}} 
\left[A\left(\frac{t+t_w}{2} \right)  + 2T \int_0^{t_w} d\tau  
f\left( \frac{t+t_w}{2} - \tau \right) g(\tau) 
\right] .
\end{equation}

The spin response to magnetic field is obtained by perturbing the Hamiltonian,
${\cal H}(t) \rightarrow {\cal H}(t) - 
\int^td\tau\sum_{\rm{x}} S_{\rm{x}}(\tau)h_{\rm{x}}(\tau)$,
with an impulse of magnetic field at time $t_w$, and is given by
\begin{eqnarray}
\label{resp-def}
R_{\rm{x}}(t,t_w) &:=& \left.
\frac{\delta \langle S_{\rm{x}+\rm{y}}(t) \rangle_{h} }
{\delta h_{\rm{y}}(t_w)}
\right|_{h=0}
\nonumber \\
&=& \frac{1}{N} \sum_k e^{i k \cdot \rm{x}} e^{- \omega_k (t-t_w)} 
\sqrt{\frac{g(t_w)}{g(t)}} ,
\end{eqnarray}
for $t \ge t_w$, and zero otherwise.
The auto-response function, $R(t,t_w):= R_{\rm{x}=0}(t,t_w)$, 
therefore takes the form
\begin{equation}
\label{autoresp}
R(t,t_w) = \sqrt{\frac{g(t_w)}{g(t)}} f\left( \frac{t-t_w}{2}\right).
\end{equation}

\subsubsection{Critical quench}
~\\
The asymptotic behaviour of $f(t) \sim \widetilde{A}_0 t^{-d/\sigma}$, and 
$A(t) \sim  \widetilde{A}_1 t^{-(d+\sigma_0)/\sigma}$, is independent of 
temperature. Making use of these asymptotics and that of $g(t)$ at $T_c$ as 
given in equation (\ref{g:T=Tc}), the expression (\ref{autocorr}) at large $t$ 
and $t_w$ reduces to
\begin{equation}
\label{autocorr1}
C(t,t_w) = 2^{d/\sigma} \widetilde{A}_0 ~t_w^{1-d/\sigma}
\left[ t_w^{-(1+ \psi+ \sigma_0/\sigma)} F_{1C}(x) + F_{2C}(x)\right],
\end{equation}
where, $x=t/t_w$ and
\begin{eqnarray}
\label{M-0}
F_{1C}(x) = \frac{\widetilde{A}_1 }{g_0 \widetilde{A}_0}2^{\sigma_0/\sigma}~ 
x^{-\psi/2} (x+1)^{-(d+\sigma_0)/\sigma}, \\
\label{K-0}
F_{2C}(x) = 2T_c ~ x^{-\psi/2} \int_0^1 \!\!dv (x+1-2v)^{-d/\sigma} v^{\psi}.
\end{eqnarray}
Thus the auto-correlation function has the scaling form
\begin{equation}
\label{autocorr2}
C(t,t_w) = 2^{d/\sigma} \widetilde{A}_0 ~t_w^{-b} ~f_C(x),
\end{equation}
with the following three possibilities:\\
(i) $b=-1+d/\sigma$ and 
$f_C(x) = F_{1C}(x) + F_{2C}(x)$,  when $1+\psi+\sigma_0/\sigma =0$;\\
(ii) $b=-1+d/\sigma$ and
$f_C(x) = F_{2C}(x) $,  when $1+\psi+\sigma_0/\sigma > 0$;\\
(iii) $b= \psi + (d+\sigma_0)/\sigma$ and $f_C= F_{1C}(x)$, when 
$1+\psi+\sigma_0/\sigma <0$.\\
\noindent
One of these possibilities occurs in each of the five regimes mentioned in 
table(\ref{table1}). In the limit $x \rightarrow \infty$
\begin{equation}
\label{f-C}
f_C(x) \sim x^{-\lambda_C/z},
\end{equation}
where the dynamic exponent $z=\sigma$. The values of $b$ and $\lambda_C$ in 
various regimes are listed explicitly in table(\ref{table2}), with one fine-tuned 
exception in regime-I. This exceptional case is when not only $\sigma_0=-\sigma$ 
but also the amplitude of the initial correlation is tuned such that it
exactly corresponds to correlations of the equilibrium long-range spherical model.
In this fine-tuned case, the system is always at equilibrium
and the value of the ageing exponent $\lambda_C = d$.

Similarly, using the asymptotics of $f(t)$ and $g(t)$ in equation (\ref{autoresp}), we 
obtain the expression for the response function for large times as
\begin{equation}
\label{autoresp1}
R(t,t_w) = 2^{d/\sigma} \widetilde{A}_0 ~
t_w^{-d/\sigma} x^{-\psi/2} (x-1)^{-d/\sigma}, 
\end{equation}
where $x \ne 1$.
Here again the auto-response function takes the expected scaling form
\begin{equation}
\label{autoresp2}
R(t,t_w) = 2^{d/\sigma} \widetilde{A}_0 ~
t_w^{-1-a} f_R(x), 
\end{equation}
where $a= -1 + d/\sigma$ and $f_R(x) = x^{-\psi/2} (x-1)^{-d/\sigma}$, which in 
the limit $x \rightarrow \infty$ behaves as
\begin{equation}
\label{f-R}
f_R(x) \sim x^{-\lambda_R/z},
\end{equation}
with $\lambda_R = d + \sigma \psi/2$. The values of $b$ and $\lambda_R$ 
in all the five regimes are listed explicitly in table(\ref{table2}).

Another interesting quantity to look at is the fluctuation-dissipation ratio 
(FDR). A deviation of this ratio from unity indicates that the state is nonequilibrium. 
FDR, which is defined as
\begin{equation}
\label{FDR-def}
X(t,t_w) :=  T \frac{R(t,t_w)} {\partial_{t_w} C(t,t_w)},
\end{equation}
can be obtained explicitly up on  
using equations (\ref{autocorr1}) and (\ref{autoresp1}), 
and is given by the expression
\begin{equation}
\label{FDR}
X(t,t_w) =  T_c \frac{x^{-\psi/2} (x-1)^{-d/\sigma}}
{ t_w^{-(1+ \psi+ \sigma_0/\sigma)} \widetilde{F}_{1C}(x) + 
\widetilde{F}_{2C}(x)} ,
\end{equation}
where, $x \ne 1$, and  
\begin{eqnarray}
\label{M}
\widetilde{F}_{1C}(x) = 
-\left(\psi+ \frac{d+\sigma_0}{\sigma}\right)F_{1C}(x) -x \frac{d}{dx} F_{1C}(x), 
\\
\label{K}
\widetilde{F}_{2C}(x) = \left(1- \frac{d}{\sigma}\right)F_{2C}(x) -
x \frac{d}{dx} F_{2C}(x).
\end{eqnarray}
The large $x$ behaviour of $\widetilde{F}_{1C}(x)$, 
using equations (\ref{M-0}) and (\ref{M}), is easily obtainable and is given by
\begin{equation}
\label{Mx}
\widetilde{F}_{1C}(x) 
\sim \frac{\widetilde{A}_1 }{g_0 \widetilde{A}_0}2^{\sigma_0/\sigma}
~ (-k_0) x^{-(k_1 + (d+\sigma_0)/\sigma)},
\end{equation}
where either $k_0=k_1=\psi/2$ when $\psi \ne 0$, or
$k_0=(d+\sigma_0)/\sigma$ and $k_1=1$ when $\psi=0$.
Similarly, the large $x$ behaviour of 
$\widetilde{F}_{2C}(x)$, 
using equations 
(\ref{K-0}) and (\ref{K}), for $\psi \ne 0$ is given by
\begin{equation}
\label{Kx}
\widetilde{F}_{2C}(x) \sim  T_c \frac{2+\psi}{1+\psi}
x^{-(\psi/2 + d/\sigma)},
\end{equation}
while for $\psi=0$,
\begin{equation}
\label{Kx-2}
\widetilde{F}_{2C}(x) =  T_c 
\left[ (x+1)^{-d/\sigma} + (x-1)^{-d/\sigma} \right].
\end{equation}
Using the above expressions for $\widetilde{F}_{1C}(x)$ 
and $\widetilde{F}_{2C}(x)$ in equation (\ref{FDR}), it
is easy to evaluate the asymptotic FDR
\begin{equation}
\label{FDR-lim}
X^\infty := \lim_{t_w \rightarrow \infty} \left( \lim_{t
\rightarrow \infty}X(t,t_w) \right) =\lim_{x\to\infty}
\left( \lim_{t_w\to\infty} \left. X(t,t_w)\right|_{x=t/s} \right).
\end{equation}
The asymptotic FDR in all cases is obtained as specified in table(\ref{table2}). 

A few remarks on the ageing exponents follow. Both regimes III and IV are 
independent of initial conditions, and the exponents are same as 
in the case of the uncorrelated initial state \cite{florian}.  
Note that in both these regimes the condition $\sigma+\sigma_0 > 0$ is 
satisfied.
A physical interpretation of this inequality is 
that the initial correlations are weaker than 
the equilibrium correlations towards which the system is evolving. 
This is easily deduced by comparing equilibrium correlations $C_{eq}(\rm{x}) \sim 
|\rm{x}|^{-(d-\sigma)}$ \cite{Joyce66} with 
initial correlations $C_{in}(\rm{x}) \sim |\rm{x}|^{-(d+\sigma_0)}$.
Also note that the auto-correlation exponent $b$ in the case of uncorrelated 
initial state \cite{florian} is not independent, since we can obtain 
$C_{eq}(\rm{x}) \sim |\rm{x}|^{-b z}$, when $b=-1+d/\sigma$ is chosen.

In regimes II and V, $\sigma+\sigma_0 < 0$, and hence can be expected that the 
ageing exponents will be influenced by the initial conditions, which indeed is the case. 

Regime I is where $\sigma+\sigma_0$ can be positive, or negative, or even zero.
Here some of the exponents, namely $a$ and $b$, are blind to initial conditions 
while others, namely $\lambda_C$ and $\lambda_R$, are not. The earlier-mentioned 
physical picture seems to fail in this regime. 
When $\sigma+\sigma_0=0$, the asymptotic FDR $X^\infty =1$, though in general 
$\lambda_C \ne \lambda_R$. The only exception is the fine-tuned case where the system 
is at equilibrium to begin with. In all the regimes which depend on initial conditions, 
$\lambda_C < \lambda_R$, suggesting that the correlations decay slower than the 
responses.

\begin{table}[t]
\[
\begin{array}{||c||c|c|c|c|c||} \hline \hline
 \mbox{Regime} & b & \lambda_C & a & \lambda_R & X^\infty\\
  \hline \hline
\mbox{I} & -1+d/\sigma  & d+(\sigma_0-\sigma)/2^* & -1+d/\sigma   
&d - (\sigma +\sigma_0)/2   &  \delta_{\sigma+\sigma_0 , 0}\\
  \hline
\mbox{II} & 1 & (d+\sigma + \sigma_0)/2 & -1+d/\sigma   
& (d + \sigma -\sigma_0)/2   & 0 \\  
\hline
\mbox{III} & -1+d/\sigma  & 3d/2 - \sigma  & -1+d/\sigma   
& 3d/2 - \sigma  & 1-\sigma/d \\
  \hline
\mbox{IV} &  -1+d/\sigma & d & -1+d/\sigma   
& d   & 1/2\\
  \hline
\mbox{V} &  (d+\sigma_0)/\sigma & d + \sigma_0 & -1+d/\sigma   
& d   & 0\\
  \hline \hline
  \end{array}
  \]
  \caption{\label{table2} Ageing exponents and asymptotic FDR in various critical 
regimes; $^*$except for the fine-tuned case, where $\lambda_C=d$}
  \end{table}

\subsubsection{Phase ordering}
~\\
We shall briefly discuss the scaling behaviour when the system evolves
from the long-range initial conditions below critical temperature. 
Substituting equation (\ref{g:T<Tc}) into expression 
(\ref{autocorr}), results in the following late-time auto-correlation function
when $T < T_c$:
\begin{equation}
\label{autocorr:P}
C(t,t_w) = \left(1-\frac{T}{T_c} \right) 
\left(\frac{(x+1)^2}{4x}\right)^{-(d+\sigma_0)/2\sigma}.
\end{equation}
Thus scaling behaviour is exhibited even while coarsening; 
with the exponents $b=0$ and $\lambda_C = (d+\sigma_0)/2$.

Similarly, substituting equation (\ref{g:T<Tc}) into expression 
(\ref{autoresp}) gives the late-time auto-response function as 
\begin{equation}
\label{autoresp:P}
R(t,t_w) = 2^{d/\sigma} \widetilde{A}_0 ~
t_w^{-d/\sigma} x^{(d+\sigma_0)/2\sigma} (x-1)^{-d/\sigma}, 
\end{equation}
and which shows scaling behaviour with the exponents
$a=-1+d/\sigma$ and $\lambda_R = (d-\sigma_0)/2$. It is easy to see, 
from equations (\ref{autocorr:P}) and ( \ref{autoresp:P}), that 
in the phase-ordering regime the asymptotic FDR vanishes, or rather 
$X(t,t_w) \to 0$ as $t_w \to \infty$ for any fixed $x$.

\section{Magnetized initial conditions}
\label{sec:mag-ini}

In this section, we evaluate the late-time auto-correlation and -response 
functions, within the Gaussian approximation, when the system evolves
at critical temperature from an initial state having a 
non-zero magnetization, 
\begin{equation}
\label{in-mag}
\frac{1}{N}\sum_{\rm{x}} \langle s_{\rm{x}}(0)\rangle = 
\frac{1}{N} \langle s_k(0)\rangle|_{k=0} = m_0 ,
\end{equation}
and uncorrelated spin-spin correlations,   
\begin{equation}
\label{in-corr}
\langle s_{\rm{x}}(0)s_{\rm{y}}(0)\rangle = (1-m^2_0) \delta_{\rm{x},\rm{y}} + m^2_0 ,
\end{equation}
which respect the spherical constraint.

Recent studies in short-range spherical model 
\cite{annibale06,annibale08} have shown that the non-Gaussian corrections are not 
relevant for local observables, and only become significant for global 
observables, when magnetized initial conditions are considered. These results 
will also hold for the long-range spherical model, with magnetized initial conditions. 

\subsection{Constraint function}

Equation (\ref{in-corr}) implies $ C_k(0,0) = (1-m^2_0) + m^2_0 N \delta_{k,0}$, 
which upon substituting into equation (\ref{A-fn}) gives 
\begin{equation}
\label{A-fn-mag}
A(t) = (1-m^2_0) f(t) + m^2_0 .
\end{equation}
Using the Laplace transform of the above function in equation (\ref{g-L}), and then
performing an inverse-Laplace transform in the small $p$ region, gives the following 
late-time behaviour of the constraint function,
\begin{equation}
\label{g-mag}
g(t) \approx g_M(\sigma, d) t^{-2\theta} \left(1 + \frac{t}{t_M} \right),
\end{equation}
where $t_M = (1-2\theta) (1-m^2_0 )/2T_c m^2_0 $, is a time scale associated with the 
initial magnetization, and
\begin{equation}
\label{theta-def}
\theta = \left\{ \begin{array}{cc}
0,~ & 0 < \sigma < d/2  \\
1-d/2\sigma,~ &  d/2 < \sigma < d  \end{array} \right. ,
\end{equation}
\begin{equation}
\label{gM-def}
g_M(\sigma, d) = \frac{1-m^2_0}{4 T_c^2} \times
\left\{ \begin{array}{cc}
1/A_2 ,~ & 0 < \sigma < d/2  \\
1/A_0 \Gamma(-1+d/\sigma) ,~ &  d/2 < \sigma < d  \end{array} \right. .
\end{equation}

\subsection{Magnetization}

We first evaluate the large-time behaviour of the magnetization, 
$m(t) := \langle s_{k=0}(t)\rangle/N $, which indeed obeys the 
expected scaling form \cite{janssen}.
Using equations (\ref{solution}), (\ref{in-mag}) and (\ref{g-mag}), 
we obtain the magnetization as
\begin{equation}
\label{mag}
m(t) = \frac{m_0}{\sqrt{g(t)}} 
\approx \frac{m_0}{\sqrt{g_M}} t^{\theta} \left(1 + \frac{t}{t_M} \right)^{-1/2}.
\end{equation}
For $t \ll t_M$, the function $m(t) \sim t^{\theta}$, and hence the initial-slip 
exponent is equal to $\theta$. 
While for $t \gg t_M$ the asymptotic behaviour of the magnetization is
$m(t) \sim t^{\theta-1/2}$. Now the large-time behaviour of the order-parameter
near the critical temperature is 
expected to relax as $t^{-\beta/\nu z}$, where $\beta$ and $\nu$ are 
the equilibrium exponents, and the dynamic exponent $z=\sigma$. For the long-range 
spherical model \cite{Joyce66} $\beta=1/2$ and 
\begin{equation}
\label{nu-data}
\nu = \left\{ \begin{array}{cc}
1/\sigma,~ & 0 < \sigma < d/2  \\
1/(d-\sigma),~ &  d/2 < \sigma < d  \end{array} \right. .
\end{equation}
Thus in this model the slip exponent is not independent  
and is given by the relation $\theta = -\beta/\nu z + 1/2$.
If we also introduce long-range correlations in the magnetized initial states, 
then $\theta$ can depend on the classes of initial correlations, 
and hence can be an independent exponent.

\subsection{Auto-correlation and auto-response functions}

Since the expectation value of the spin is non-zero in the presence of initial 
magnetization, the Fourier transform of the connected correlation is given by
\begin{equation} 
\label{con-corr}
\widetilde{C}_k(t,t_w) = C_k(t,t_w) - N \delta_{k,0} m(t) m(t_w).
\end{equation} 
Therefore, auto-correlation function, $\widetilde{C}(t,t_w):= 
\widetilde{C}_{\rm{x}=0}(t,t_w) = \sum_k \widetilde{C}_k(t,t_w) /N $, is given by
\begin{equation}
\label{autocorr-mag}
\widetilde{C}(t,t_w) = \frac{1}{\sqrt{g(t)g(t_w)}} 
\left[(1-m^2_0)f\left(\frac{t+t_w}{2} \right)  + 2T_c \int_0^{t_w} \!\!d\tau  
f\left( \frac{t+t_w}{2} - \tau \right) g(\tau) 
\right] .
\end{equation}
Using the asymptotic form of $g(t)$, as given in equation (\ref{g-mag}), and 
$f(t) \sim \widetilde{A}_0 t^{-d/\sigma}$, and the fact that the above defined 
$\theta < 1/2$, equation (\ref{autocorr-mag}) reduces to 
\begin{equation}
\label{autocorr1-mag}
C(t,t_w) \approx \widetilde{A}_0 2^{d/\sigma}  ~t_w^{1-d/\sigma} F_C(x,u),
\end{equation}
where
\begin{equation}
\label{FC-def}
F_C(x,u) = \frac{2T_c ~ x^{\theta}}{\sqrt{(1+ux)(1+u)}} 
\int_0^1 dv (x+1-2v)^{-d/\sigma} v^{-2\theta} (1+uv) ,
\end{equation}
$x=t/t_w > 1$, and $u=t_w/t_M$. Hence the value of the exponent $b=-1+d/\sigma$.
In the ageing regime, $x \gg 1$, the above function takes the asymptotic form 
\begin{equation}
\label{FC-asym}
F_C(x,u) \sim
 \frac{2T_c ~ x^{\theta-d/\sigma}}{\sqrt{(1+ux)(1+u)}} 
\left(\frac{1}{1-2\theta} +u \frac{1}{2-2\theta} \right), 
\end{equation}
and has ageing exponent $\lambda_C$, as listed in the table(\ref{table3}),
which depends on whether $t_w \ll t_M$ or $t_w \gg t_M$.

The auto-response function for the case with non-zero initial magnetization, 
from equations (\ref{autoresp}) and (\ref{g-mag}), is obtained as
\begin{equation}
\label{autoresp-mag}
R(t,t_w) = 2^{d/\sigma} \widetilde{A}_0 ~ t_w^{-d/\sigma} F_R(x,u),
\end{equation}
where
\begin{equation}
\label{FR-def}
F_R(x,u) = \sqrt{ \frac{1+u}{1+ux} } ~(x-1)^{-d/\sigma} x^{\theta}.
\end{equation}
Hence the value of the exponent $a=-1+d/\sigma$.
The ageing exponent $\lambda_R$, which again depends on whether $t_w \ll t_M$ or  
$t_w \gg t_M$, is given in table(\ref{table3}). 
Both the exponents, $a$ and $b$, do not notice the time-scale $t_M$ and
are unaffected by the initial magnetization.

The expression for FDR, obtained from equations (\ref{autocorr1-mag}) and 
(\ref{autoresp-mag}), reduces to
\begin{equation}
\label{FDR-mag}
X(t,t_w) = \frac {T_c F_R(x,u)}
{\left( 1-d/\sigma + u \partial_u - x \partial_x \right)F_C(x,u) } , 
\end{equation}   
and results in the following asymptotic value of FDR,
\begin{equation}
\label{FDRval-mag}
X^{\infty} = \left\{ \begin{array}{cc}
(1-2\theta)/(2-2\theta) ,~ &  t_w \ll t_M  \\
2(1-\theta)/(3-2\theta),~ &  t_w \gg t_M 
\end{array} \right. ,
\end{equation}   
which is explicitly given in the table(\ref{table3}).

In the case where $t_w \ll t_M$, the initial magnetization is irrelevant and the
ageing exponents are the same as those in the case of the uncorrelated initial state with no 
initial magnetization. As the waiting time $t_w$ increases and becomes larger than 
$t_M$, the system will crossover to a new magnetized ageing class.

\begin{table}[t]
\[
\begin{array}{||c||c|c|c||} \hline \hline
 \mbox{Conditions} & a=b &\lambda_C = \lambda_R & X^\infty\\
  \hline \hline
t_w \ll t_M ,~ 0<\sigma< d/2   & -1+d/\sigma  &d   &  1/2 \\
  \hline
t_w \ll t_M ,~ d/2<\sigma< d &  -1+d/\sigma  & 3d/2 - \sigma    & 1-d/\sigma \\  
\hline
t_w \gg t_M ,~ 0<\sigma< d/2   & -1+d/\sigma & d + \sigma/2  & 2/3 \\
  \hline
t_w \gg t_M ,~ d/2<\sigma< d & -1+d/\sigma   & (3d-\sigma)/2   & d/(d+\sigma)\\
  \hline \hline
  \end{array}
  \]
  \caption{\label{table3} Ageing exponents and asymptotic FDR  
in the Gaussian regime in presence of initial magnetization. }
  \end{table}

\section{Conclusion}
\label{conc}
To summarize, the correlation and response functions for the non-conserved long-range 
spherical model 
are determined for two different sets of initial conditions:\\
(i) long-range initial correlations with an arbitrary power $\sigma_0 > -d$, and 
vanishing magnetization; \\
(ii) non-zero initial magnetization $m_0$, but no initial correlations beyond spherical 
fluctuations. \\
We evaluated the ageing exponents corresponding to various ageing classes. 
The ageing exponents can depend on the class of the initial state, though not on the 
details of it. 

When the long-range initial conditions are considered, the correlations and responses 
do show scaling behaviour for all $\sigma_0 > -d$, in both the mean-field and the 
non-trivial cases of the long-range spherical model.

In the mean-field case ($ 0 < \sigma < d/2$), as the value of $\sigma_0$ increases we 
distinguish three regimes: \\
(i) regime II, where $-d < \sigma_0 < \sigma-d$, \\
(ii) regime V, where $\sigma_0 > \sigma-d$  and $\sigma_0 < -\sigma$, \\ 
(iii) regime IV, where $\sigma_0 > \sigma-d$ and $\sigma_0 > -\sigma$. \\
In both regimes II and V, the initial correlations are stronger than the equilibrium 
correlations, while in regime IV it is the other way around. We find that
the exponents in regimes II and V can depend on $\sigma_0$, whereas not in regime IV.
In other words, we find that as we reduce the initial correlations there is a crossover 
from the initial-condition dependent ageing classes to that with no dependence. 

In the non-trivial case ($ d/2 < \sigma < d$), as the value of $\sigma_0$ increases we 
distinguish two regimes: \\
(i) Regime I, where $-d < \sigma_0 < \sigma-d$, \\
(i) Regime III, where $\sigma_0 > \sigma-d$.  \\
We find that the exponents in regime I can depend on $\sigma_0$, whereas this is not the case 
in regime III. Here too there is a crossover from the initial-condition dependent ageing 
class to that with no dependence, as $\sigma_0$ goes from below to above $\sigma-d$.
In regime III, the initial correlations are weaker than the equilibrium correlations.
While in regime I, depending on the value of $\sigma$, it can be weaker, similar, or 
stronger, though the exponents depend of $\sigma_0$. We lack a simple physical explanation of 
this dependence on the initial conditions in regime I.

In the case of non-vanishing initial magnetization too, the correlations and 
responses show scaling behaviour. The initial magnetization $m_0$ introduces a 
time scale $t_M$, and when the waiting time $t_w \ll t_M$, the ageing classes are same 
as those obtained with $m_0=0$. As $t_w$ increases beyond $t_M$, there is a cross over 
to the magnetized ageing classes.

Though we have evaluated only auto-correlation and auto-response functions, a similar 
analysis will suggest that the form of the two-point functions, in the presence of 
initial correlations, gets modified from that given in equation (\ref{two-pt-fn}) to
the following form,
\begin{eqnarray}
\label{two-pt-fn2}
\left\langle {\cal O}_i(\rm{x},t) {\cal O}_j(\rm{y}, t_w) \right\rangle = 
t_w^{-a_{ij}} F_{ij}\left(\frac{|\rm{x}-\rm{y}|}{t_w^{1/z}}, \frac{t}{t_w} \right)
+t_w^{-\widetilde{a}_{ij}} 
\widetilde{F}_{ij}\left(\frac{|\rm{x}-\rm{y}|}{t_w^{1/z}}, \frac{t}{t_w} \right),
\end{eqnarray}
where the second term on the right carries the memory of initial conditions,
and defines new exponent $\widetilde{a}_{ij}$ and scaling function $\widetilde{F}_{ij}$.
The function asymptotically behaves as 
$\widetilde{F}_{ij}(0,x) \sim x^{-\widetilde{\lambda}_{ij}/z}$, and defines
another exponent $\widetilde{\lambda}_{ij}$.
When $\widetilde{a}_{ij} \le a_{ij}$, then the fluctuations at time $t$
within regions of length-scale much less than $t^{1/z}$ are no longer canonical 
equilibrium fluctuations. The above scaling form for the two-point functions is likely 
to hold beyond the model that we considered here.

~\\
\noindent
{\bf Acknowledgements:}
I thank Hyunggyu Park, and the statistical physics group at LPM, Nancy,
for useful discussions. This work was done with the generous support 
by the people of South-Korea, including that by the Korea Foundation for 
International Cooperation of Science and Technology(KICOS) through a 
grant provided by the Korean Ministry of Science and Technology(MOST) 
with No. 2007-00369.


\begin{thebibliography}{99}
\bibitem{hohenberg}
P. C. Hohenberg and B. I. Halperin, Rev. Mod. Phys. {\bf 49}, 435 (1977).

\bibitem{bray94}
A. J. Bray, Adv. Phys. {\bf 43}, 357 (1994).

\bibitem{calabrese05} 
P. Calabrese and A. Gambassi, J. Phys. A, {\bf 38}, R133, (2005).

\bibitem{florian}
F. Baumann, S. B. Dutta, and M. Henkel, 
J. Phys. A: Math. Theor. {\bf 40}, 7389 (2007).

\bibitem{janssen}
H. K. Janssen, B. Schaub, and B. Schmittmann, Z. Phys. B{\bf 73}, 539 (1989).

\bibitem{bray91}
A. J. Bray, K. Humayun, and T. J. Newman, Phys. Rev. B {\bf 43}, 3699 (1991).

\bibitem{ritschel}
U. Ritschel and H. W. Diehl, Phys. Rev. E {\bf 51}, 5392 (1995).

\bibitem{picone}
A. Picone and M. Henkel, J. Phys. A: Math. Gen. {\bf 35}, 5575 (2002).

\bibitem{fedorenko} 
A. A. Fedorenko and S. Trimper, Europhys. Lett. {\bf 74}, 89 (2006). 

\bibitem{annibale06} 
A. Annibale and P. Sollich, J. Phys. A: Math. Gen. {\bf 39}, 2853 (2006).

\bibitem{calabrese06} 
P. Calabrese, A. Gambassi and F. Krzakala, J. Stat. Mech. P06016  (2006).

\bibitem{calabrese07} 
P. Calabrese and A. Gambassi, J. Stat. Mech. P01001 (2007).  

\bibitem{annibale08} 
A. Annibale and P. Sollich,  J. Phys. A: Math. Theor. {\bf 41}, 135001 (2008).

\bibitem{Godreche00} C. Godr{\`e}che and J-M. Luck, J. Phys. A:
Math. Gen. {\bf 33}, 9141 (2000).

\bibitem{Joyce66} 
G.S. Joyce in C. Domb and M. Green (eds) {\it Phase transitions
and critical phenomena}, Vol. 2, ch. 10, London (Academic 1972).

\bibitem{fusco}
N. Fusco and M. Zannetti, Phys. Rev. E {\bf 66}, 066113 (2002).

\end{thebibliography}
\end{document}